\documentclass[runningheads]{llncs}
\usepackage[T1]{fontenc}
\usepackage{graphicx}
\usepackage{amsmath}
\usepackage{subfigure}
\usepackage{textcomp}
\usepackage{url}
\usepackage{tikz}
\usepackage{xcolor}
\usepackage{algorithm2e}
\usepackage{booktabs}
\usepackage{hyperref}
\hypersetup{hidelinks} 
\makeatletter
\g@addto@macro\UrlBreaks{\do\-}
\makeatother
\definecolor{myblue}{RGB}{0,112,192}
\newcommand*\circled[1]{\tikz[baseline=(char.base)]{
		\node[shape=circle,draw=myblue,inner sep=1pt, fill=myblue, text=white] (char) {#1};}}
\newcommand{\head}[1]{\noindent\textbf{#1}}

\begin{document}
\title{VR-Themis: A Scalable Framework for Virtual Reality Application Clone Detection}
\titlerunning{VR-Themis: VR Application Clone Detection}
\author{
Gengyang Xu\inst{1,4} \and
Hanyang Guo\inst{2} \and
Hong-Ning Dai\inst{1}\thanks{Corresponding author.} \and
Weizhi Meng\inst{3}
}

\authorrunning{G. Xu et al.}

\institute{
Hong Kong Baptist University, Hong Kong, China\\
\email{henrydai@comp.hkbu.edu.hk}
\and
Sun Yat-Sen University, Guangdong, China
\and
Lancaster University, Lancaster, UK
\and
Hong Kong University of Science and Technology, Hong Kong, China\\
\email{gxuah@cse.ust.hk}
}
\maketitle              
\begin{abstract}
Repackaging of mobile applications (aka app cloning) not only threatens the security and privacy of mobile users but also infringes upon the copyright of the original app developers.
However, existing detection methods that primarily focus on mobile platforms (such as Android) fail to capture the essential features of virtual reality (VR). Consequently, they are inadequate for effectively detecting cloned VR apps, which have often been targeted by illegal users in the VR market.
Considering the unique features of VR apps, this paper proposes a two-stage app clone detection framework, namely VR-Themis, based on \emph{Hierarchy-Object-Behaviour} (HOB). Firstly, VR-Themis exploits the coarse-grained stage to cluster apps based on their retrievable statistical features, making this tool scalable to large-scale VR app datasets. Then, in the fine-grained stage, VR-Themis performs in-depth analysis of the suspicious apps (identified in the first stage) by calculating similarity using our defined \emph{HOB metrics}. Our extensive experiments indicate that VR-Themis successfully detects 307 suspected clone apps from the collected 4,277 VR apps without false positives, demonstrating its effectiveness and scalability.

\end{abstract}
\section{Introduction}
Repackaged VR apps pose a significant threat to VR ecosystems. They undermine the interests of original developers, increase the maintenance burden on app markets, and threaten user privacy. For instance, Commuter Games reported that their VR racing game, \textit{Downtown Club}, has about five times more players than the copies sold. Similarly, Realcast's \textit{Just Hoops} has nearly six times as many users as copies sold~\cite{Bezmalinovic2024}. This phenomenon is largely due to the presence of pirated versions repackaged~\cite{Bezmalinovic2024}. Furthermore, repackaged VR apps significantly increase users' vulnerability to security and privacy attacks (e.g., inception attacks~\cite{Yang2024}). Although some VR headset companies have released solutions (e.g., Meta Quest Attestation API~\cite{attestationAPI}), their reliance on online connectivity makes them vulnerable to crafty circumvention. As of this writing, active VR piracy communities continue to thrive, disseminating sources for repackaged pirated app downloads and methods for installing unauthorized apps on VR headsets, posing a lasting challenge to the industry~\cite{Reddit2025a,Reddit2025b,Reddit2023,VRPirates2023}.

In fact, the VR ecosystem is not the first mobile ecosystem to face the threat of app cloning. The ease of repackaging on mobile platforms has made them hotspots for illegal users. For example, Android apps are easily repackaged using reverse engineering tools, such as Apktool~\cite{Apktool2016}, dex2jar~\cite{dex2jar2011}, and JADX~\cite{jadx2013}. These tools enable users to decompile apps, modify intrinsic contents, and then repackage them for distribution.
Existing state-of-the-art clone detectors for mobile apps rely on various features for similarity comparison, including code-centric similarity~\cite{DNADroid,DroidMOSS}, layout similarity of User Interface (UI)~\cite{DroidEagle}, resource metadata comparison~\cite{FSquaDRA}, and combinations of them~\cite{ViewDroid,ResDroid}.

Despite the numerous existing clone detection approaches for mobile apps, VR platforms introduce novel development features that render conventional methods entirely ineffective for detecting clones in VR applications. 
Traditional mobile app development primarily relies on tools like Android Studio~\cite{AndroidStudio2023}, focusing on \textit{2D} interface design and application logic. As a result, existing mobile app clone detectors typically use code, UI layouts, and resource metadata as the main features in their similarity metrics. 
In contrast, the development of VR apps emphasizes \textit{immersive 3D scene} design. According to a breakdown of VR development costs, primary budget allocations include 30\% for digital assets (e.g., \textit{3D} modeling and animation), 25\% for code development, and 15\% for \textit{3D} user interaction design~\cite{Wrexa2024}. This allocation reflects the increased importance of non-code asset creation and \textit{3D} scene design in VR, highlighting the limitations of mobile-oriented clone detection methods. In summary, existing state-of-the-art mobile app clone detection approaches are inherently infeasible for VR apps due to three key reasons:
(1) The \textit{3D} scene structures of VR apps cannot be represented by the \textit{2D} UI layouts used in mobile applications; (2) VR apps heavily rely on non-code assets (e.g., \textit{3D} models and animations) that are not considered in mobile-oriented detection methods; and (3) \textit{3D} development engines (e.g., Unity) typically provide protection mechanisms (e.g., IL2CPP in Unity), which invalidate code-centric similarity metrics. These disparities \textit{necessitate a novel detection approach tailored specifically for VR ecosystems}, rather than incremental improvements over existing mobile-oriented methods.

To bridge this research gap, this paper proposes a novel clone detection approach specifically designed for VR apps, named VR-Themis\footnote{The name VR-Themis is inspired by the Greek goddess Themis, who can distinguish faked/cloned objects.}. 
This detection approach adopts a two-stage framework and is specifically designed to address three critical challenges in VR app clone detection: 

(1) \textbf{Complexity of Pairwise Comparison.} Directly comparing all possible pairs of VR apps becomes computationally prohibitive with large datasets. VR-Themis addresses this challenge through a coarse-grained stage that groups VR apps based on retrievable statistical features, thereby significantly reducing the complexity of comparisons. 

(2) \textbf{Limitations of Conventional Clone Detection Characteristics and Metrics.}
Existing mobile app clone detection methods rely on features such as code, UI layouts, and resource metadata, which are ineffective for the unique complexities of VR apps. Unlike mobile apps, VR apps involve intricate \textit{3D} scene hierarchies with interconnected \texttt{\small GameObject}s, immersive visual elements, dynamic interaction components, and customized script Behaviours.
To characterize these unique features of VR apps, we propose the Hierarchy-Object-Behaviour (HOB) model, a structured representation tailored for VR. Additionally, conventional similarity metrics fall short in capturing the multi-dimensional complexities of VR apps. To overcome this challenge, VR-Themis introduces \emph{HOB metrics}, a three-tiered similarity metric suite based on the \emph{HOB} model by integrating \emph{Hierarchy Edit Distance} for structural differences, \emph{GameObject Node Distance} for visual and functional variations, and \emph{Script Behaviour Similarity} for Behavioural consistency. This synergy allows VR-Themis to perform fine-grained and multi-dimensional comparisons of suspicious cloned apps.

(3) \textbf{Scarcity of Benchmark Datasets for VR Clone Detection:} Unlike the mobile app ecosystem, the VR domain lacks publicly available datasets specifically designed for proprietary VR app clone detection. To the best of our knowledge, there is currently no benchmark tailored for proprietary VR app clones with the provision of reliable ground truths. Additionally, existing VR-related datasets from other research do not meet the scale requirement for our study. Consequently, we constructed our own dataset consisting of 4,277 VR apps from multiple sources (more details in~\textsection\ref{sec:dataset}). For our experiments, we also utilize a smaller, carefully curated subset of this dataset, with manually verified clone truths, to determine thresholds and report accuracy.

In summary, the main contributions of this work are highlighted as follows:
\begin{itemize}
\item We construct a VR app dataset collected from diverse sources, including side-loading channels that have not been considered in previous research. 

\item  We propose an automatic app clone detection approach. As far as we know, this is the first study on proprietary VR app clone detection.
Our two-stage method incorporates a coarse-grained clustering to reduce comparison space, followed by fine-grained in-depth comparisons.

\item We introduce \emph{HOB metrics}, a VR similarity metric suite based on our \emph{HOB} model. This metric suite is specifically designed to compare the unique characteristics of VR apps, including intricate \textit{3D} scene hierarchies, object-level visual and functional components, and script-driven Behaviours.

\item We conduct extensive experiments on 4,277 Unity-based VR apps. Results show that VR-Themis effectively and efficiently detects cloned VR apps with high accuracy. Our artifact is available at~\cite{vr-themis}.
\end{itemize}

\section{Background and Threat Model}
\subsection{Mobile Application Clone}
App cloning has become a significant concern in the mobile app industry, as it often facilitates various cybercrimes. Common motivations include: inserting malicious code into popular apps~\cite{Li2017}, hijacking advertising revenue~\cite{DroidMOSS}, distributing cracked versions that bypass paid features~\cite{Reddit2023}, creating unauthorized localizations~\cite{VRmoo2025}, and plagiarizing proprietary \textit{3D} assets with minimal effort~\cite{Zuo2023}. The last motivation is particularly disruptive in VR, where high-quality \textit{3D} models and animations require substantial development resources.

\subsection{VR Application Development}
VR operating systems (OSs) (e.g., Meta Horizon OS~\cite{Meta_Horizon_OS} and PICO OS~\cite{PICO_OS}) are custom-developed on top of Android OS. However, VR application development differs significantly from conventional mobile app development (e.g., Android SDK) due to the unique requirements of VR platforms.
VR development emphasizes \emph{immersion} that enables users to engage naturally with their surroundings. Unlike traditional mobile app development, where coding plays a predominant role, VR development places significant importance on creating high-quality \textit{3D} assets and designing immersive \textit{3D} scenes.

A VR app typically consists of several scenes, where a scene is defined as an asset that contains all or part of an app~\cite{UnityDoc2022}. The VR scene representation serves as the structural foundation for organizing the visual, functional, and interactive elements within \textit{3D} virtual environments. In most VR development engines, this representation adopts a tree structure, commonly referred to as the \emph{hierarchy}. The \emph{hierarchy} represents the grouping and parent-child relationships among \texttt{\small GameObject}s~\cite{Takoordyal2020}. Each node in the scene's hierarchy tree corresponds to a \texttt{\small GameObject}, where the parent \texttt{\small GameObject} may contain other child \texttt{\small GameObject}s inheriting the parent's properties. 
\texttt{\small GameObject}s are the fundamental elements in the VR engine to represent real objects like balls, lights, cameras, and so on~\cite{UnityDoc2022}. Nevertheless, \texttt{\small GameObject}s cannot perform any functions themselves; they only serve as containers. To enable a \texttt{\small GameObject} to possess a specific object's necessary properties, one or more components must be attached to it. 
In Unity, various built-in components are provided, though developers can also create their own components using the Unity Scripting API~\cite{Takoordyal2020}.

\subsection{Threat Model} 
We define the threat model from the following perspectives.

    \textbf{Goal:} The goal of VR-Themis is to accurately identify \textit{cloned VR apps}, which are defined as two (or more) apps exhibiting substantial similarity in architecture and fundamental assets but are distributed under distinct ownership without explicit authorization. This focus distinguishes VR app clone detection from code reuse detection, as the latter does not emphasize scene hierarchies and digital assets.
    
    \textbf{Adversary capabilities:} An adversary possesses the following capabilities:
    \begin{itemize}
        \item \emph{Decompilation:} The adversary can use tools like AssetStudio~\cite{AssetStudio2022} to decompile VR APKs and extract scene hierarchies, assets (e.g., meshes, textures, and animations), and C\# scripts.
        \item \emph{Modification:} The adversary can obfuscate or minify scripts (e.g., renaming methods and variables in C\#), modify \textit{3D} models (e.g., squashing, stretching, and vertex perturbation), and alter scene hierarchies.
        \item \emph{Distribution:} The adversary can distribute cloned apps through unofficial channels (e.g., sideloading platforms) or masquerade as developers.
    \end{itemize}
    
    \textbf{Assumptions:} 
    First, we assume that most VR apps store their scene hierarchies and assets statically within the VR APK. This assumption is supported by the fact that the majority of VR apps are designed to function in an offline way. Apps that dynamically load resources at runtime or retrieve assets from remote servers are outside the scope of this work. 
    Second, we assume that the adversary is capable of performing minor modifications to cloned VR apps to circumvent detection. These modifications may include code obfuscation, asset perturbation, and alterations to the scene hierarchy. However, the adversary is unlikely to fully redesign the app or make significant architectural changes due to resource and time constraints.

\section{Hierarchy-Object-Behaviour Similarity Metric for VR App Clone Detection}
This section introduces a modeling framework for VR app clone detection by combining a representation with its similarity metrics. We first present the \emph{Hierarchy-Object-Behaviour (HOB)} model, which encodes a VR app as scene hierarchies of \texttt{\small GameObject}s, comprising objects with components and script-driven Behaviours. Building on this representation, we define \emph{HOB metrics}, a multi-dimensional suite of similarity measures. Together, the \emph{HOB} model and \emph{HOB metrics} capture VR-specific characteristics that extend beyond mobile-oriented features. We depict the \emph{HOB} model, formalize \emph{HOB metrics}, and explain their integration into VR-Themis's fine-grained comparison as follows.

\begin{figure}[t] 
    \centering 
    \includegraphics[width=0.5\textwidth]{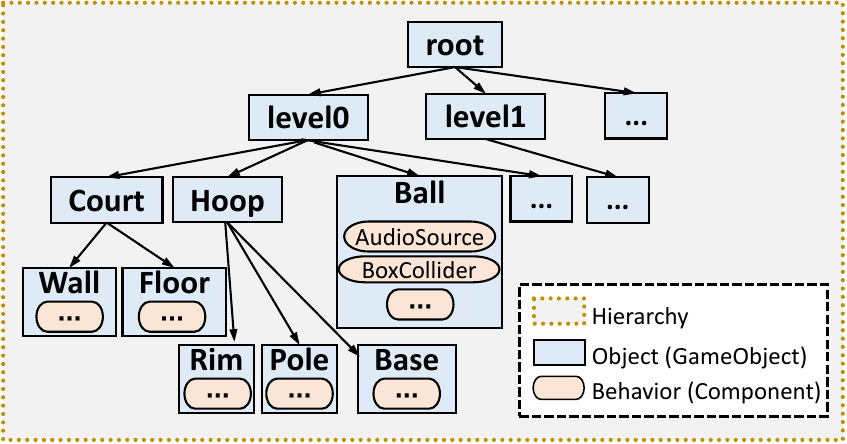} 
    \caption{An example of a VR app represented by the \emph{HOB} model} 
    \label{fig:Example_HOB} 
\end{figure}

\subsection{Hierarchy-Object-Behaviour Model} 
To overcome the limitations of conventional mobile-oriented app clone detectors in capturing VR-specific characteristics, we propose a HOB model. This model provides a structured representation of VR apps by analyzing their \textit{scene hierarchy (Hierarchy)}, \textit{3D objects (Object)}, and \textit{script-driven components (Behaviour)}. Fig.~\ref{fig:Example_HOB} illustrates an example of a VR application represented by the \emph{HOB} model. By comprehensively capturing the characteristics of VR apps from high-level structural representations to detailed functional Behaviours, it effectively overcomes the shortcomings of mobile app clone detection methods in handling \textit{3D} scenes and non-code assets.

\head{Hierarchy.}
VR apps typically consist of multiple scenes (levels), each represented as a hierarchy tree. This structure describes the parent-child relationships and organization of all objects (i.e., \texttt{\small GameObject}) within a scene. The HOB model parses and extracts this hierarchical information from VR APK, organizing the VR app as a tree structure, in which each scene is represented as a sub-tree. Each node of a tree corresponds to a \texttt{\small GameObject} containing the following information: 
\textit{Components} are multiple attachments defining the properties and Behaviours of the \texttt{\small GameObject}; and \textit{Transform} includes \textit{3D} coordinates, rotation, and scale. In this way, the HOB model comprehensively reflects the architecture of a VR app.

\head{Object.}
\label{sec:HOB-Object}
In the HOB model, the definition of an object is based on the \texttt{\small GameObject} concept. \texttt{\small GameObject}'s functionality is enabled through the attachment of components. However, game engines lack an effective mechanism to classify \texttt{\small GameObject}s based on their \emph{functionality}. This limitation introduces significant challenges when comparing \texttt{\small GameObject}s across different VR apps, as it requires examining all their components individually. Such an approach is both computationally expensive and inefficient, especially given the diversity and complexity of components.
To address this issue, we introduce a classification mechanism that assigns a \emph{type} to each \texttt{\small GameObject} based on its major components. This mechanism groups \texttt{\small GameObject}s with similar functionalities under the same \emph{type}.
For instance, when an \texttt{\small AudioSource}\footnote{The \texttt{\footnotesize AudioSource} component endows the \texttt{\footnotesize GameObject} with the ability to play sounds in a \textit{3D} environment.} component is attached to a \texttt{\small GameObject}, we classify this \texttt{\small GameObject} as an Audio type. Our proposed \texttt{\small GameObject} \emph{type} classification mechanism adheres to three criteria: 1) \emph{Comprehensiveness} requires defining as many \texttt{\small GameObject} types as possible to ensure broad coverage; 2) \emph{Accuracy} ensures that the assigned \emph{type} accurately represents the \texttt{\small GameObject}'s functionality; and 3) \emph{No overlapping of type definitions} guarantees that a \texttt{\small GameObject} cannot belong to multiple \emph{type}s simultaneously.
To ensure accurate classification, we implement a priority mechanism that assigns each \texttt{\small GameObject} a unique \emph{type} based on its highest-priority component, avoiding overlaps. 

\head{Behaviour.}
In VR apps, \textit{Behaviours} are the core components that determine the interactivity and functionality, and their definition is based on the concept of \texttt{\small Component}. Behaviours are typically implemented through C\# scripts, which extend the \texttt{\small MonoBehaviour} base class to define the functionality of \texttt{\small GameObject}. For example, Behaviours can trigger events, modify component properties, or respond to user inputs. Unlike built-in components provided by game engines, Behaviours consist of \emph{custom code written by developers}, often using tools like the Unity Scripting API~\cite{Takoordyal2020}. Illegal cloners frequently make minor modifications to Behaviour scripts, such as renaming variables, to evade detection. However, since Behaviour scripts directly control the functionality of \texttt{\small GameObject}s, their core logic is difficult to alter. Therefore, by extracting the key features of Behaviours, it is possible to identify Behavioural similarities for VR clone detection.

\subsection{HOB Metrics: Multi-Dimensional VR App Similarity Measures}

Effectively detecting VR app clones requires a reliable similarity metric that can comprehensively capture the structural, object-level (visual and functional), and Behavioural characteristics of VR applications. To this end, we propose a three-tiered \textit{HOB metrics} suite specifically tailored for VR applications based on our HOB model. \textit{HOB metrics} include three core components: \textit{Hierarchy Edit Distance (HED)} evaluates the structural differences between two VR scene hierarchy trees; \textit{GameObject Node Distance (GND)} quantifies object-level visual and functional differences; and \textit{Script Behaviour Similarity (SBS)} assesses the similarity of script-driven Behaviours within \texttt{\small GameObject}s. Together, these metrics provide a comprehensive framework for capturing high-level structural transformations, detailed object-level variations, and subtle Behaviour-level modifications.

\head{HED} measures differences between two VR scene hierarchy trees. 
HED is defined as the minimum cost of transforming one scene hierarchy tree into another through a series of operations.
Let \( \mathcal{T} \) denote the collection of VR app scene hierarchy trees, represented as \( \mathcal{T} = \{ T_1, T_2, \ldots, T_n \} \), where \( n \) indicates the total number of apps within the collection. Each tree \( T_i \) ($i\in[1,2,\ldots,n]$) consists of a collection of nodes, denoted by set \( N_i \), where each node is indexed as \( n_{i,j} \in N_i \) for \( j \in [1, 2, \ldots, m] \), with \( m \) being the total number of nodes in tree \( T_i \). 
We mainly consider the following operations involved in transforming a scene hierarchy tree: insertion, deletion, substitution, and modification on a \texttt{\small GameObject}. The difference between substitution and modification lies in the former replacing a \texttt{\small GameObject} with a different \emph{type}, while the latter modifies the components of a \texttt{\small GameObject} without changing its type. The cost of each operation is within the range \([0, 1]\). Specifically, the costs for node deletion, insertion, and substitution operations are all equal to \(1\) while the cost of node modification is equal to the distance between two \texttt{\small GameObject}s, to be measured by GND (see below). 

\RestyleAlgo{ruled}

\begin{algorithm}[t]
    \caption{GameObject Node Distance (GND) calculation}
    \label{alg:gnd_calculation}
    \small
\DontPrintSemicolon
\KwIn{Two \texttt{\footnotesize GameObject}s $G_k$ and $G_l$.}
\KwOut{Node distance $\delta_\text{GND}$.}
\SetKwProg{Fn}{Function}{}{}
\Fn{\upshape GNDCalculation($G_k$, $G_l$)}{
  \If{$G_k$.\texttt{\upshape type} $\neq G_l$.\texttt{\upshape type}}{
    \Return 1
  }
  \If{$G_k$.\texttt{\upshape type} = \texttt{\upshape ``3DObject''}}{
    \If{$G_k$.\texttt{\upshape getMeshHash()} $\neq G_l$.\texttt{\upshape getMeshHash()}}{
      \Return 1
    }
  }
  intersectionSet $\gets \emptyset$\;
  unionSet $\gets \emptyset$\;
  \ForEach{$c_k \in G_k.\texttt{\upshape components}$}{
    unionSet.add$(c_k.\text{type})$\;
    \ForEach{$c_l \in G_l.\texttt{\upshape components}$}{
      \If{$c_k.\text{type} = c_l.\texttt{\upshape type}$}{
        \eIf{$c_k.\text{type} = \texttt{\upshape ``MonoBehaviour''}$}{
          similarity $\gets \textsc{ComputeSBS}(c_k, c_l)$\;
          \If{$\texttt{\upshape similarity} > 0.8$}{
            intersectionSet.add$(c_k.\text{type})$\;
          }
        }{
          intersectionSet.add$(c_k.\text{type})$\;
        }
      }
    }
  }
  unionSet.add$(G_l.\text{components})$\;
  $\delta_\text{GND} \gets 1 - \dfrac{|\operatorname{intersectionSet}|}{|\operatorname{unionSet}|}$\;
  \Return $\delta_\text{GND}$\;
}
\end{algorithm}

\begin{figure}[t]
    \centering
    \subfigure[Scene hierarchy tree $T_1$.\label{fig:sceneHierarchyTreeStructure_T1}]{
        \includegraphics[width=0.48\textwidth]{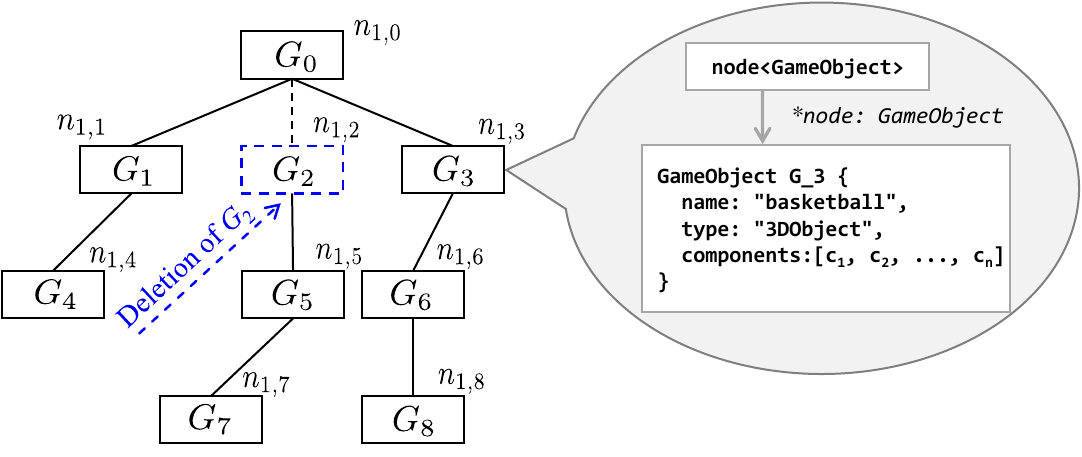}}
    \hfill
    \subfigure[Scene hierarchy tree $T_2$.\label{fig:sceneHierarchyTreeStructure_T2}]{
        \includegraphics[width=0.48\textwidth]{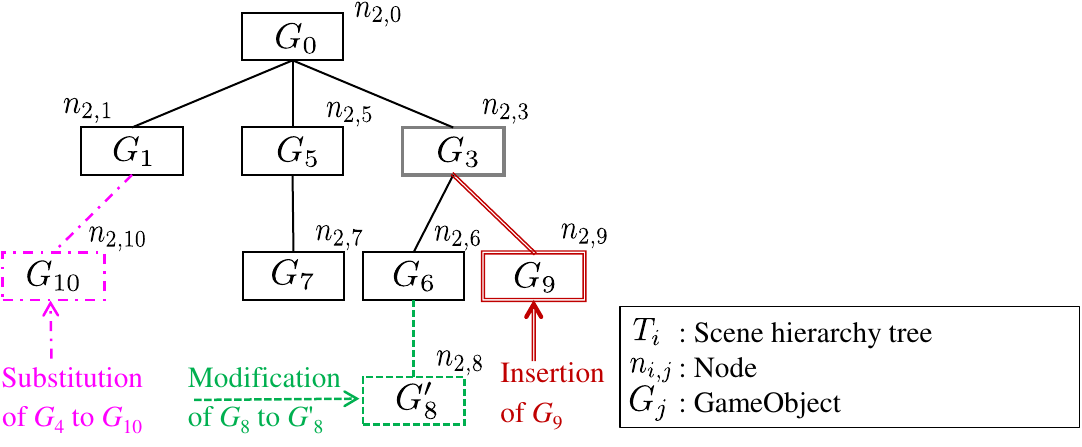}}
    \caption{Examples of two scene hierarchy trees and their HED}
    \label{fig:sceneHierarchyTreeStructure}
\end{figure}

\head{GND} denoted by $\delta_\text{GND}$ quantifies the cost of modifying a \texttt{\small GameObject} by analyzing its components. Each \texttt{\small GameObject} is characterized by a set of various components, such as \texttt{\small MeshRenderer}, \texttt{\small AudioSource}, and \texttt{\small MonoBehaviour}. For most \texttt{\small GameObject} types, GND reflects functional differences by evaluating the Jaccard distance between two \texttt{\small GameObject}s' component sets. Let \( G_k \) and \( G_l \) denote two \texttt{\small GameObject}s being compared. Their component sets are represented as \( C_k = G_k.\text{components} \) and \( C_l = G_l.\text{components} \), respectively. Then, $\delta_\text{GND}$ is expressed as:
\begin{equation}
\delta_\text{GND}(G_k, G_l) = J(C_k, C_l) = 1 - \frac{|C_k \cap C_l|}{|C_k  \cup  C_l|},    
\label{eq:gnd}
\end{equation}
where $J(\cdot,\cdot)$ denotes the Jaccard distance. 
However, an exception occurs when two \texttt{\small GameObject}s of type ``\texttt{\small 3DObject}'' have different meshes. For \texttt{\small 3DObject} nodes, whose primary feature is their \textit{visual representation} rather than functional differences, GND adopts a specialized \textit{mesh comparison method} (to be elaborated in \textsection\ref{sec:fineStage}). If the meshes are different, their \texttt{\small GameObject} node distance is directly assigned to \(1\) without further comparison. This is because different meshes result in completely different visual information, making them incomparable even if all other components are identical. Algorithm \ref{alg:gnd_calculation} outlines the process for calculating the GND. 
Fig.~\ref{fig:sceneHierarchyTreeStructure} depicts an example of measuring two trees, \( T_1 \) and \( T_2 \) by HED. Each node is linked to a corresponding \texttt{\small GameObject}. For example, node \( n_{2,3} \) corresponds to \texttt{\small GameObject} \( G_3 \) in Fig. \ref{fig:sceneHierarchyTreeStructure}, in which node insertion and deletion only have a cost of 1, while the cost of modification is determined by Eq.~\eqref{eq:gnd}. 

\head{SBS} measures similarity between scripts. When comparing two components, it is often sufficient to simply compare their types (built-in by the game engine), as this approach adequately reflects their functionality. However, for custom-defined script components, it is necessary to compare the corresponding scripts to assess their Behaviours. To address this issue, we define SBS, which employs cosine similarity to evaluate the feature vectors extracted from scripts (details in \textsection\ref{sec:fineStage}). For two \texttt{\small MonoBehaviour} components \( c_k \) and \( c_l \), we denote their script feature vectors by \( f_k = \text{fingerprint}(c_k) \) and \( f_l = \text{fingerprint}(c_l) \), respectively. The cosine similarity, denoted by $\gamma$, is defined as follows:
\begin{equation}
\gamma = \cos(\alpha) = \frac{f_k \cdot f_l}{\|f_k\| \|f_l\|} = \frac{\sum_{u=1}^{d} f_{ku} \cdot f_{lu}}{\sqrt{\sum_{u=1}^{d} f_{ku}^2} \times \sqrt{\sum_{u=1}^{d} f_{lu}^2}},
\end{equation}
where \( \alpha \) denotes the angle between two vectors, and \( d \) is the dimension of the script feature vectors. When their cosine similarity is greater than or equal to \(0.8\), the components can be deemed identical. This threshold is chosen based on the common practice in mobile app repackaging detection, where sharing 80\% of the code serves as a typical indicator~\cite{Li2019}.

\section{Design of VR-Themis}
\begin{figure*} [t]
       \centering 
       \includegraphics[width=\textwidth]{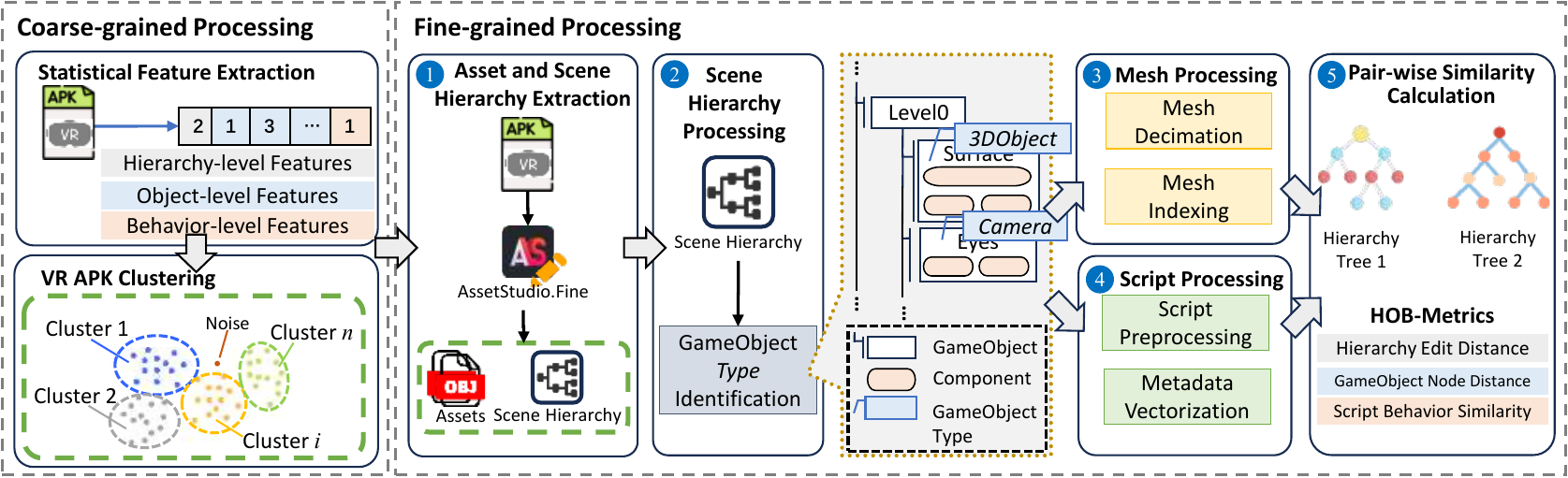} 
       \caption{Overview of VR-Themis} 
       \label{fig:VR-Themis-Framework} 
   \end{figure*}

This section elaborates on the design of VR-Themis, illustrated in Fig.~\ref{fig:VR-Themis-Framework}, covering its scope and two-stage workflow: coarse-grained and fine-grained processing.

\subsection{Scope}
VR-Themis is specifically designed to detect cloned VR apps. This study primarily focuses on VR apps developed for Meta Quest series headsets due to two key reasons: 1) Investigating other Android-based VR platforms may introduce complications due to the same app being distributed across multiple device stores. Verifying whether detected suspected clones arise from legitimate cross-platform distribution or unauthorized cloning requires substantial additional effort, incurring unnecessary complexity. 2) Meta Quest series dominates the VR headset market~\cite{Armstrong2023}, providing a large and diverse dataset for our investigation. Nonetheless, our detection framework is flexible and can be seamlessly adapted to other Android-based VR platforms, such as Pico~\cite{PICO_OS} and Vive~\cite{HTC2025}. 

Furthermore, this study concentrates on VR apps developed by the Unity engine for the following two reasons: 1) Over 70\% of top-selling Quest games have been developed on top of Unity~\cite{UnityVR2024}, making it a natural and significant focus for our study; 2) To extract VR assets, we utilize existing GUI-based, engine-specific asset extraction tools (e.g., AssetStudio~\cite{AssetStudio2022} for Unity apps and UE Viewer for Unreal Engine) and adapt them for automated asset extraction.

\subsection{Coarse-grained Processing}
\label{sec:coarseStage}
The primary goal of the coarse-grained stage is to reduce the comparison complexity (space) since directly comparing all possible pairs of VR apps becomes computationally intensive with large datasets. In this stage, VR-Themis clusters collected VR apps based on statistical features specifically designed for VR. Grounded in our HOB model, we selected ten key features covering structural, object-level (visual and functional), and Behavioural aspects of VR applications. The feature selection process was based on an in-depth analysis of VR development practices. It was validated by three experienced VR developers, each with more than three years of industry expertise. This process ensures the features are both representative and comprehensive in characterizing VR apps.

Specifically, the selected features are elaborated at three different levels.  The hierarchy level includes (1) the number of \texttt{\small levels} (scenes). At the object level, features encompass (2) the number of \texttt{\small GameObject} nodes, (3) \texttt{\small AnimationClips}, (4) \texttt{\small Animators}, (5) \texttt{\small Materials}, (6) \texttt{\small Sprites}, (7) \texttt{\small Texture2Ds}, (8) \texttt{\small AudioClips}, and (9) \texttt{\small meshes}. The Behaviour level includes (10) the number of \texttt{\small MonoBehaviours}. The imbalanced number of features across different levels is explained as follows. The object level is designed to capture the visual and functional characteristics of VR. Each type of object, such as meshes and animations, provides distinct information, necessitating multiple features for accurate clustering. 
In contrast, the hierarchy level and Behaviour-level scripts are difficult to distinguish using statistical features alone. 
Therefore, they need more detailed analysis at the fine-grained stage, using \emph{HED} for structural comparisons and \emph{SBS} for Behaviour assessments.

To automate feature extraction, we modified the original \texttt{\small AssetStudio}~\cite{AssetStudio2022} and developed a console-based version to automatically extract and statistically analyze asset features.
After extracting the key features of VR apps, we apply a grouping process to categorize VR apps with similar attributes into the same group based on their feature fingerprints (details in \textsection\ref{sec:evaluation-coarse}).

\subsection{Fine-grained Processing}
\label{sec:fineStage}
After clustering VR apps with similar features, we proceed to the fine-grained stage, where a detailed comparison of VR apps is conducted based on our HOB model. This stage comprehensively evaluates VR apps by analyzing their structural, object-level, and Behavioural similarities using the \emph{HOB metrics} framework.
In Step~\circled{\scriptsize 1}, we extract the VR scene hierarchy and corresponding VR assets. In Step~\circled{\scriptsize 2}, we organize a scene hierarchy tree for each VR application, which serves as the basis for calculating the HED. Nodes containing meshes and scripts undergo further analysis in Step~\circled{\scriptsize 3} (Mesh Processing) and Step~\circled{\scriptsize 4} (Script Processing). At the object level, GND quantifies visual and functional variations by comparing the components of GameObjects. At the Behavioural level, SBS evaluates the consistency of custom scripts by analyzing their extracted feature vectors. Finally, in Step~\circled{\scriptsize 5}, we calculate pairwise similarity scores for VR apps by integrating HED, GND, and SBS into \emph{HOB metrics}. 

\head{Asset and Scene Hierarchy Extraction.}
To extract the VR scene hierarchy and corresponding VR assets (e.g., meshes), we implement a custom tool based on AssetStudio~\cite{AssetStudio2022}. While AssetStudio offers general asset extraction capabilities, it does not sufficiently support the extraction of scene hierarchies and VR-specific assets (e.g., environment and navigation elements). To address these limitations, we developed a specialized tool called \texttt{\small AssetStudio.Fine}, which is tailored to account for the unique characteristics of Unity-developed VR apps.

\head{Scene Hierarchy Processing.}
To facilitate the pairwise comparison of VR apps, we first organize the scene hierarchy tree of each app into a structured format, which serves as the foundation for calculating the HED. In a VR app, the \texttt{\small GameObject}s within the scene hierarchy serve as containers that aggregate various functional components. As discussed in \textsection\ref{sec:HOB-Object}, instead of exhaustively comparing all components within two \texttt{\small GameObject}s to calculate GND, which is computationally prohibitive, VR-Themis streamlines this process by categorizing \texttt{\small GameObject}s into predefined \emph{type}s based on their key components. Further component-level comparisons are conducted only when two \texttt{\small GameObject}s belong to the same \emph{type}.

To identify the components within \texttt{\small GameObject}s, VR-Themis leverages Unity's YAML Class ID Reference~\cite{UnityYAML2022}, which provides a detailed mapping of built-in Unity components to their respective class IDs. This enables VR-Themis to efficiently parse Unity asset files and accurately determine component types.

In summary, we implement a new tool (i.e., \texttt{\small AssetStudio.Fine}) based on AssetStudio to address the above challenges. This tool can extract both the scene hierarchy and assets, with each \texttt{\small GameObject} node in the hierarchy tree containing the \texttt{\small GameObject} \emph{type} and corresponding components. Representative \texttt{\small GameObject} types include \texttt{\small 3DObject}, \texttt{\small Lighting}, \texttt{\small Environment}, and so on. For \texttt{\small GameObject}s containing specific components, such as \texttt{\small MeshFilter} and \texttt{\small MonoBehaviour}, further comparisons of meshes and scripts are conducted in Step~\circled{\scriptsize 3} and Step~\circled{\scriptsize 4}, respectively.

   \begin{figure*}[t] 
       \centering 
       \includegraphics[width=0.9\textwidth]{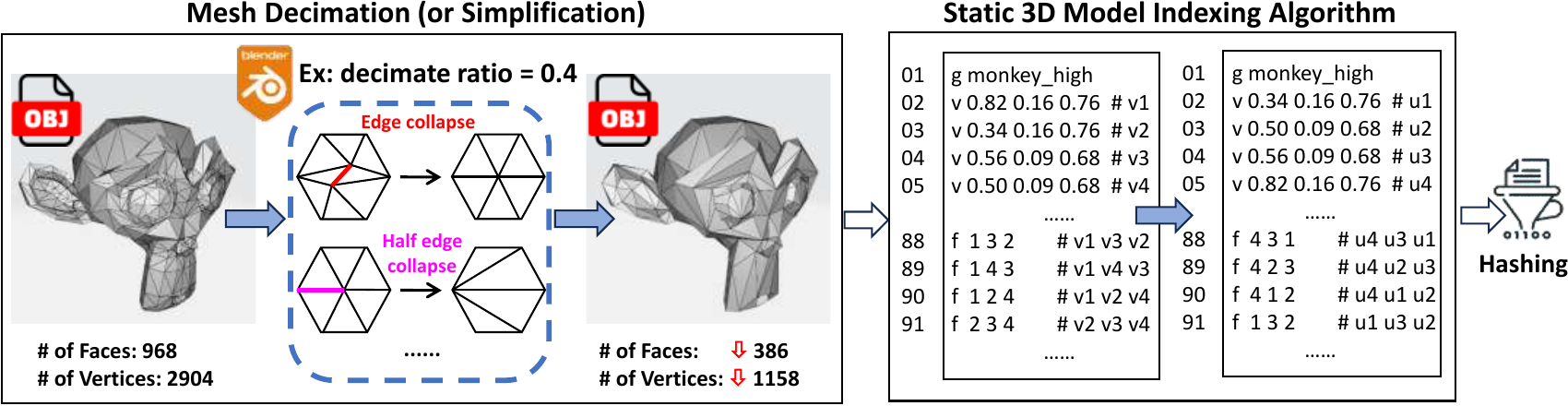} 
       \caption{Mesh processing pipeline} 
       \label{fig:MeshProcessingPipeline} 
   \end{figure*}

\head{Mesh Processing.}
As defined in GND, after identifying the \texttt{\small{GameObject}} node in the scene hierarchy tree in Step~\circled{\scriptsize 2}, we proceed to compare whether the identified node of \texttt{\small{3DObject}} type shares the same mesh. The primary goal of this step is to efficiently compare meshes (i.e., \texttt{\small MeshFilter}) contained by two \texttt{\small GameObject}s. As far as we know, only one recent study, \textsc{3DScan}~\cite{Zuo2023}, meets our requirements. In a nutshell, \textsc{3DScan} first transforms \textit{3D} objects into a number of faces, each of which is represented by vertices and edges. It then normalizes the \textit{3D} objects by de-duplication and sorting faces. Thereafter, \textsc{3DScan} compares two \textit{3D} objects by calculating their hash values of faces to represent meshes. Despite the advent of \textsc{3DScan}, it cannot efficiently handle our large experimental scale, as it takes a considerable amount of time to process complex \textit{3D} objects with a high number of faces and vertices. To tackle this challenge, we propose a novel method by adopting the \emph{mesh decimation} approach into the mesh processing algorithm. As shown in Fig.~\ref{fig:MeshProcessingPipeline}, mesh decimation can reduce the number of vertices and faces of a mesh while minimizing shape changes~\cite{Kobbelt1998}. We implement this new design by using Blender~\cite{Blender1995}, which is an open-source \textit{3D} graphics software. \textsection\ref{sec:evaluation-fine} will present the experimental results, highlighting the effectiveness of our approach and its improvements over \textsc{3DScan}.

\head{Script Processing.}
During the Unity-based VR application development, the C\# scripts can be used to trigger events, modify component properties, and respond to user input. To simplify development, all scripts created by developers in Unity are derived from the \texttt{\small MonoBehaviour} base class by default~\cite{UnityDoc2022}. The \texttt{\small MonoBehaviour} class provides a skeleton to attach scripts to \texttt{\small GameObject}s and offers hooks for common events, such as \texttt{\small Start()} and \texttt{\small Update()}. When using \texttt{\small AssetStudio.Fine} to identify the component as \texttt{\small MonoBehaviour}, we can determine that it is a custom-defined component, thereby enabling us to retrieve the corresponding C\# class (script) from the associated \texttt{\small MonoScript}.

When comparing the \texttt{\small MonoBehaviour} components of two \texttt{\small GameObject}s, a thorough comparison of the corresponding C\# scripts is necessary to ascertain whether they are indeed the same component. Therefore, we extract and decompile the code from the VR APK by reverse engineering, though the completeness of source code decompilation largely depends on the scripting backend used. Currently, Unity provides two scripting backends: 1) \emph{Mono-based decompilation}, 
2) \emph{Intermediate Language to C++} ({IL2CPP}). 
Compared with the former one, it is more difficult to decompile IL2CPP-based APKs to source code. 

\begin{figure*}[t] 
       \centering 
       \includegraphics[width=0.9\textwidth]{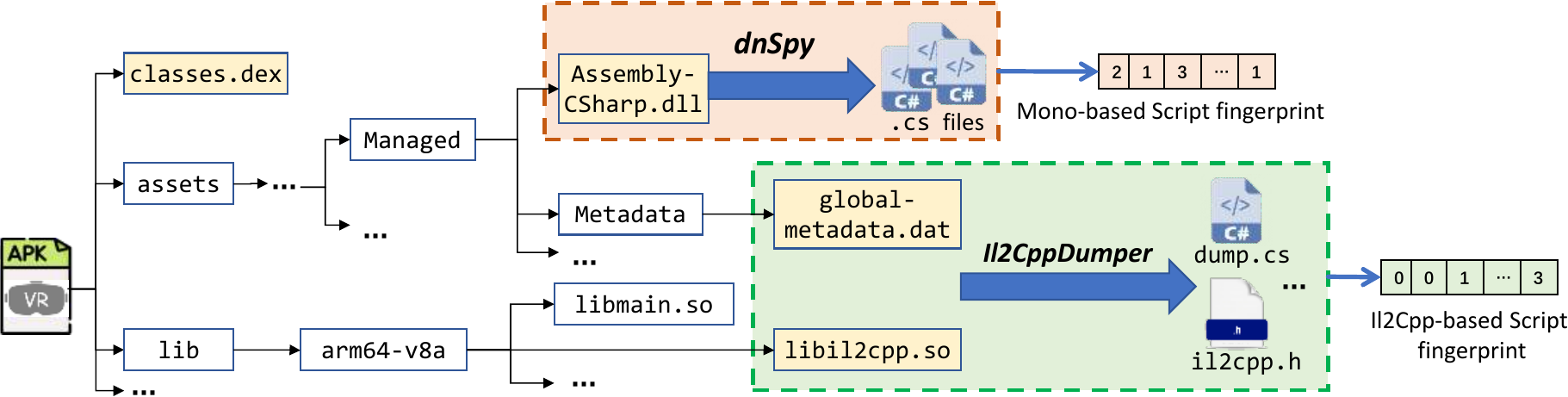} 
       \caption{C\# script decompilation under the two Unity scripting backends} 
       \label{fig:scriptDecompile} 
\end{figure*}

To ensure compatibility across different Unity scripting backends and address the challenges of decompiling source code from IL2CPP-based APKs, VR-Themis utilizes the SBS metric to evaluate script Behavioural similarity. SBS calculates the cosine similarity between feature vectors extracted from scripts, with each feature vector constructed based on key script metadata such as declarations of methods (signatures) and data attributes. This allows VR-Themis to identify Behavioural similarities even when scripts have undergone lightweight modifications, such as variable renaming or minor structural changes.
First, we need to extract the script metadata by decompilation. With regard to two different scripting backends, we have two different processing methods: 1) for Mono-based VR apps, we extract the compiled logic code file \texttt{\small Assembly-CSharp.dll} and utilize the reverse engineering tool \textit{dnSpy}~\cite{dnSpy2020} to retrieve the C\# source code; 2) for IL2CPP-based VR apps, we first extract the compiled binary files \texttt{\small libil2cpp.so} as well as the function mapping file \texttt{\small global-metadata.dat} from the VR APK. We then employ the reverse engineering tool \textit{Il2CppDumper}~\cite{Il2CppDumper2017} to perform the DLL restoration, consequently obtaining the script metadata. 
To prevent attackers from obfuscating method (function) names, we compare the modifiers, parameters, and data types (e.g., \texttt{\small \textcolor{blue}{private} List<EventTrigger.Entry>}, \texttt{\small \textcolor{blue}{public} Coroutine (IEnumerator routine)}) by removing names. We then order these features and count their quantities to construct a feature vector, which represents the script fingerprint.
Fig.~\ref{fig:scriptDecompile} depicts how the above process works.

\head{Pairwise Similarity Calculation.}
To evaluate the similarity between two VR apps, we analyze their scene hierarchies, which are fundamentally determined by their \texttt{\small GameObject} nodes and associated components. Leveraging our proposed \emph{HOB metrics}, tailored for \emph{VR scene hierarchy similarity}, we comprehensively capture structural, object-level (functional and visual), and Behavioural differences. These metrics seamlessly integrate HED, GND, and SBS to provide a robust multi-dimensional similarity assessment.

\begin{figure*}[t] 
       \centering 
       \includegraphics[width=1.0\textwidth]{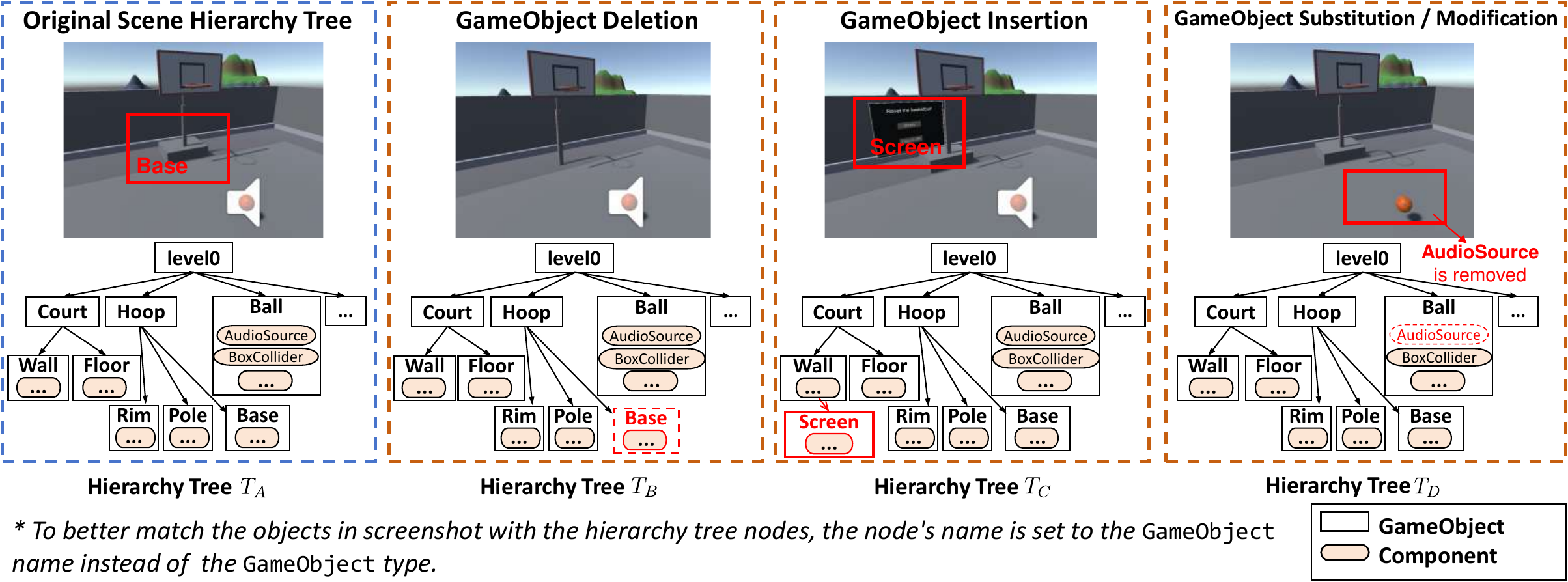} 
       \caption{Scene hierarchy tree transformation operations in an example VR scene} 
       \label{fig:sceneShots} 
\end{figure*}

In the following, we present an example to demonstrate how \emph{HOB metrics} are calculated on a real VR scene.
Fig.~\ref{fig:sceneShots} illustrates the transformation operations occurring in a real VR scene (excerpted from a real VR app). For example, a deletion operation removes the ``Base'' \texttt{\small GameObject} node from Hierarchy Tree $T_A$, resulting in Hierarchy Tree $T_B$, as shown in the screenshot (highlighted in a red box). In another example, an insertion operation adds the ``Screen'' \texttt{\small GameObject} node (highlighted in a red box) to Hierarchy Tree $T_A$, producing Hierarchy Tree $T_C$. Moreover, modifying the ``Ball'' \texttt{\small GameObject} node in Hierarchy Tree $T_A$ by removing the ``\texttt{\small AudioSource}'' component results in Hierarchy Tree $T_D$. As shown in examples of Fig.~\ref{fig:sceneShots}, the HED value between Hierarchy Trees $T_A$ and $T_B$ (or $T_C$) is \(1\), while the HED value between Hierarchy Trees $T_A$ and $T_D$ is \(0.25\) according to Eq.~\eqref{eq:gnd} since the original ``Ball'' \texttt{\small GameObject} contains four components.

\section{Implementation and Evaluation}
We evaluate VR-Themis on 4,277 VR apps. All experiments are conducted on a workstation with an Intel i7-13700 CPU and 32.0 GB of memory.

\subsection{Dataset}
\label{sec:dataset}
\begin{figure}[t]
    \centering
    \begin{minipage}[t]{0.48\textwidth}
        \centering
        \includegraphics[width=0.95\textwidth]{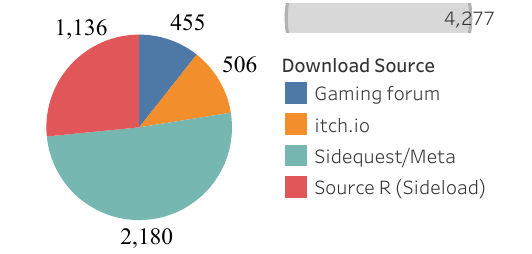}
        \caption{Distribution of APK sources}
        \label{fig:APK_source}
    \end{minipage}
    \hfill
    \begin{minipage}[t]{0.48\textwidth}
        \centering
        \includegraphics[width=0.95\textwidth]{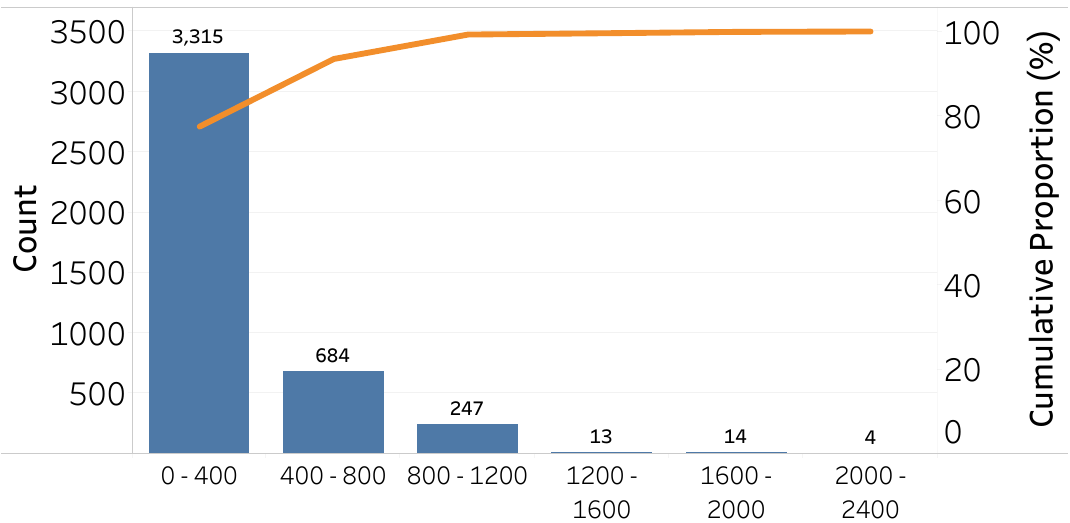}
        \caption{Distribution of APK sizes (MB)}
        \label{fig:APK_Size}
    \end{minipage}
\end{figure}

We have collected a total of 4,277 Unity-based Meta Quest VR APK files from five distinct sources: 1) the official Meta app store~\cite{MetaAppStore2023}, 2) the SideQuest platform~\cite{SideQuest2021}, 3) the \emph{itch.io} website~\cite{itchio2013}, 4) Gaming forums, and 5) Source R (i.e., a sideloading platform)\footnote{We have chosen not to disclose sideloading sources' names to prevent attracting more people to these non-compliant websites that offer unlicensed apps.} (the complete app list is available in our repository). Due to the close collaboration between the Meta app store and SideQuest~\cite{Shane2024}, the two platforms share a significant number of VR apps. Therefore, we do not explicitly differentiate between them to avoid confusion over app ownership. 
We analyze the distribution of collected apps, shown in Fig.~\ref{fig:APK_source}. For a detailed analysis, Fig.~\ref{fig:APK_Size} presents the size distribution of the APKs, which range from 18.26\,MB to 2305.95\,MB, with most being \textless 400\,MB.
To the best of our knowledge, this is the largest VR APK dataset constructed for clone detection to date. 
In comparison, existing VR APK datasets include 1,096 VR apps~\cite{Zhan2024}, 408 VR/AR apps~\cite{Alghamdi2025}, 500 Oculus Quest apps~\cite{Guo2024}, and 900 Meta Quest apps~\cite{Guo2025MetaVR}.

We collect the VR APKs through two distinct approaches. For those sources that only support the direct installation of apps onto VR headsets, we installed VR apps onto a Meta Quest 2 VR headset and extracted the corresponding APK files from the device. For others, we implement a web crawler to download the collected VR APK files. 

\subsection{Coarse-grained Detection}
\label{sec:evaluation-coarse}
Direct pairwise comparisons of all possible VR app pairs in a large dataset are computationally prohibitive. To address this issue, we adopt a coarse-grained clustering stage that significantly reduces the comparison space by grouping apps with similar features. As discussed in \textsection\ref{sec:coarseStage}, each VR APK is represented by a feature fingerprint comprising ten statistical characteristics grounded in our HOB model. In the coarse-grained stage, we employ the Density-Based Spatial Clustering of Applications with Noise (DBSCAN) algorithm~\cite{10.1145/3068335} for clustering, as it is particularly well-suited to our task for several reasons. First, unlike $K$-means, DBSCAN does not require specifying the number of clusters (\(k\)) a priori, which is advantageous given the unknown distribution of the dataset. Second, DBSCAN is robust to noise (e.g., non-cloned apps) and capable of discovering clusters with arbitrary shape (i.e., capturing diverse cloning strategies). Finally, DBSCAN's computational complexity 
ensures scalability to large datasets, whereas hierarchical clustering and spectral clustering are computationally intensive.

\begin{figure}[t]
    \centering
    \begin{minipage}[t]{0.48\textwidth}
        \centering
        \includegraphics[width=0.95\textwidth]{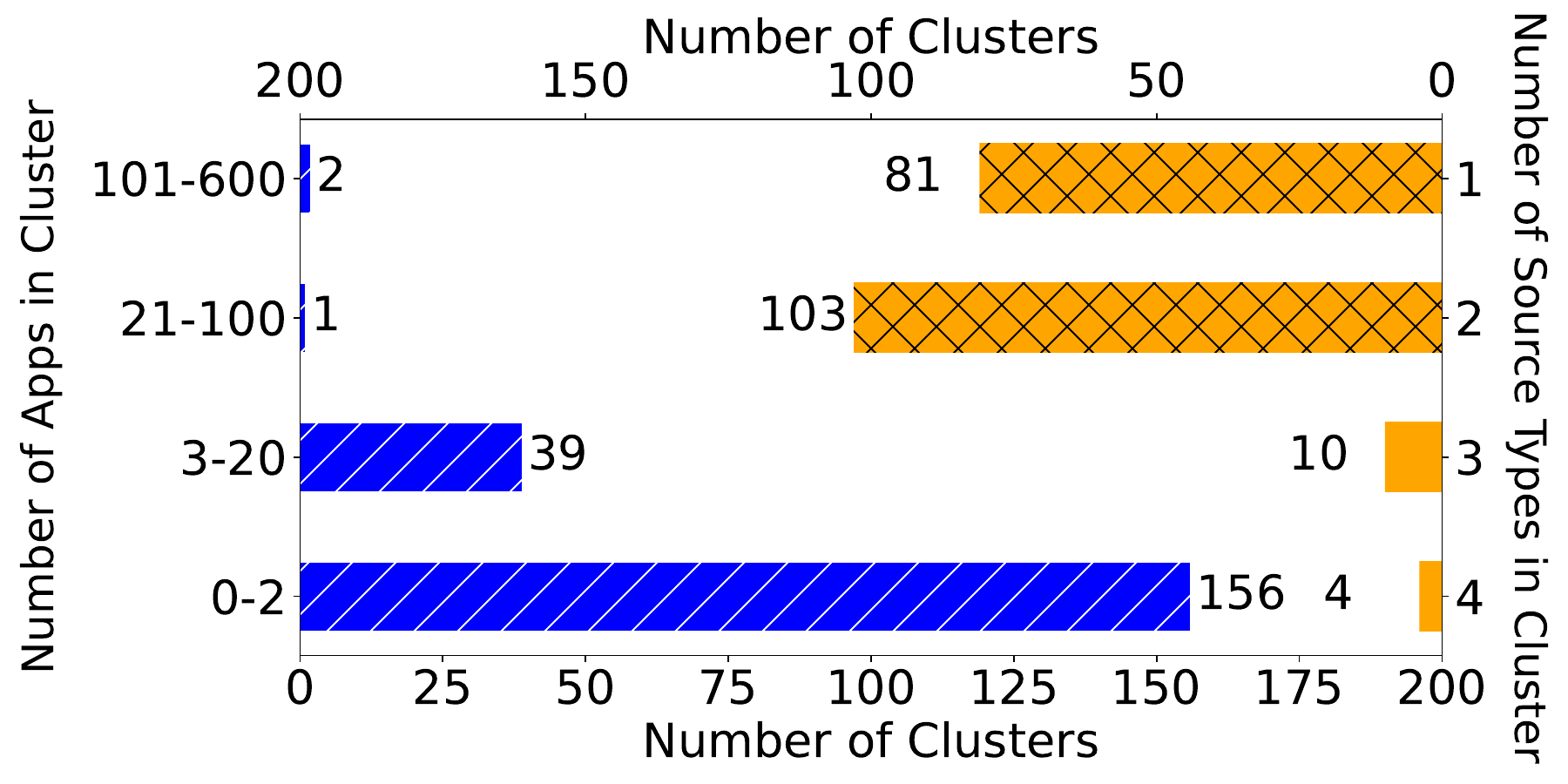}
        \caption{Distributions of cluster sizes and of app sources within clusters}
        \label{fig:clusterSize}
    \end{minipage}
    \hfill
    \begin{minipage}[t]{0.48\textwidth}
        \centering
        \includegraphics[width=0.85\textwidth]{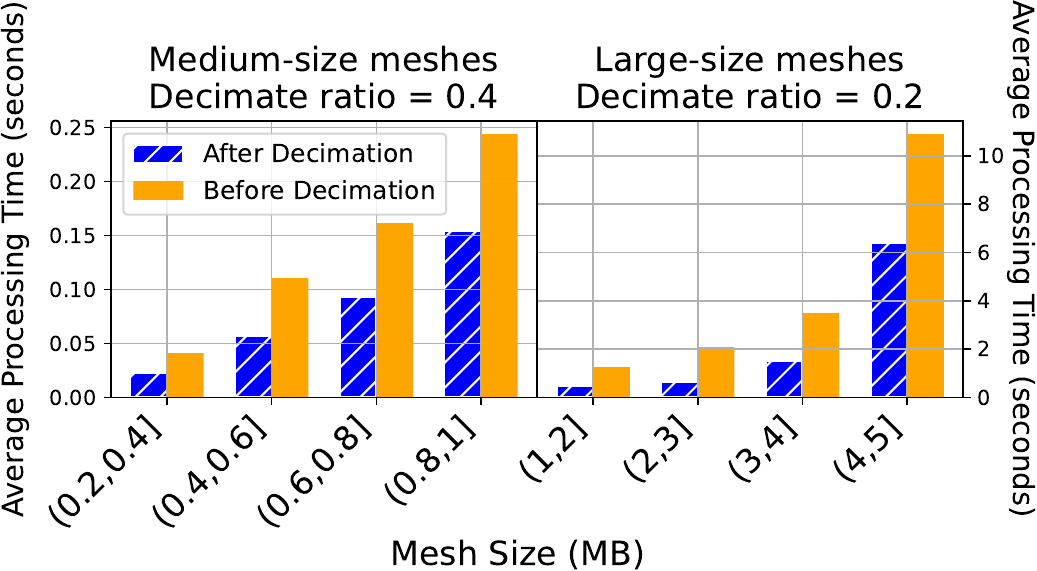}
        \caption{Average mesh processing time before and after \emph{mesh decimation}}
        \label{fig:MeshProcessingTime}
    \end{minipage}
\end{figure}

To determine DBSCAN's parameters, we set \textit{minPts} = 2 based on domain knowledge, as clone detection inherently involves at least two apps forming a potential clone relationship. For the \(\epsilon\) parameter (neighborhood radius), we used a $k$-distance graph to identify the ``elbow'' point, narrowing \(\epsilon\) to the range \((0, 1.0]\). We then randomly select 100 samples from the 4,277 apps and manually install and execute them to determine the clone truths. Validation experiments on this subset of 100 VR apps with known clone truths confirmed that \(\epsilon = 0.1\) maximized recall (100\%) while maintaining high precision (80.65\%). Feature extraction for all 4,277 VR apps required 5,810.44 seconds (approximately 1.36 seconds per app), and DBSCAN clustering completed in just 0.1024 seconds, producing 198 clusters. The distribution of cluster sizes and app sources in clusters is shown in Fig.~\ref{fig:clusterSize}, in which most clusters contain fewer than 20 apps. Furthermore, most apps within a cluster originate from one or two different sources. Finally, we reduced the initial 9,144,226 app pairs to 416,385 (about 4.55\%) suspected clone pairs, substantially lowering the computational burden for the fine-grained stage.

\begin{figure}[b]
    \centering
    \begin{minipage}[t]{0.48\textwidth}
        \centering
        \includegraphics[width=0.95\textwidth]{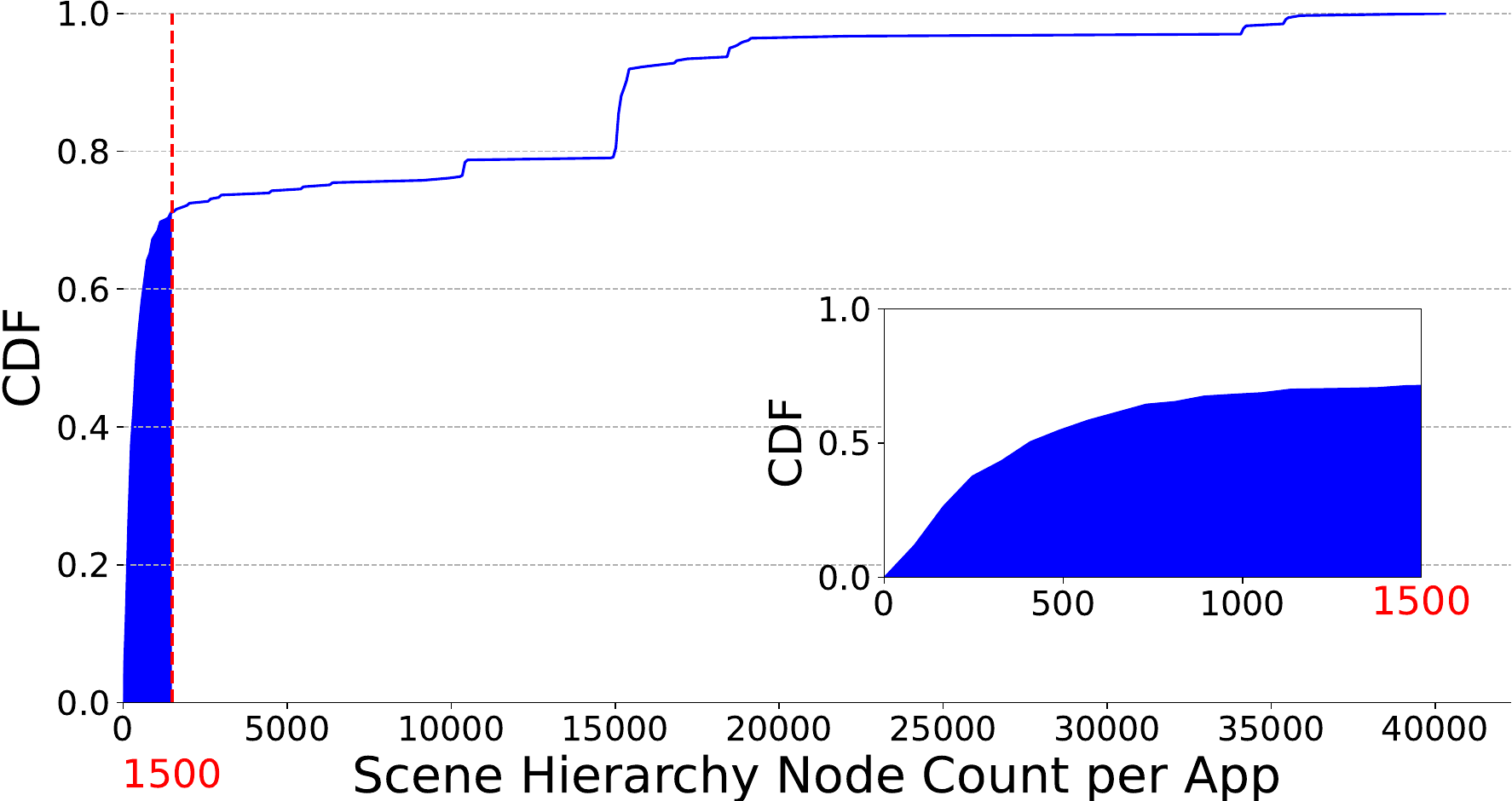}
        \caption{Distribution of node counts in app scene hierarchy trees}
        \label{fig:NodeCounts}
    \end{minipage}
    \hfill
    \begin{minipage}[t]{0.48\textwidth}
        \centering
        \includegraphics[width=0.95\textwidth]{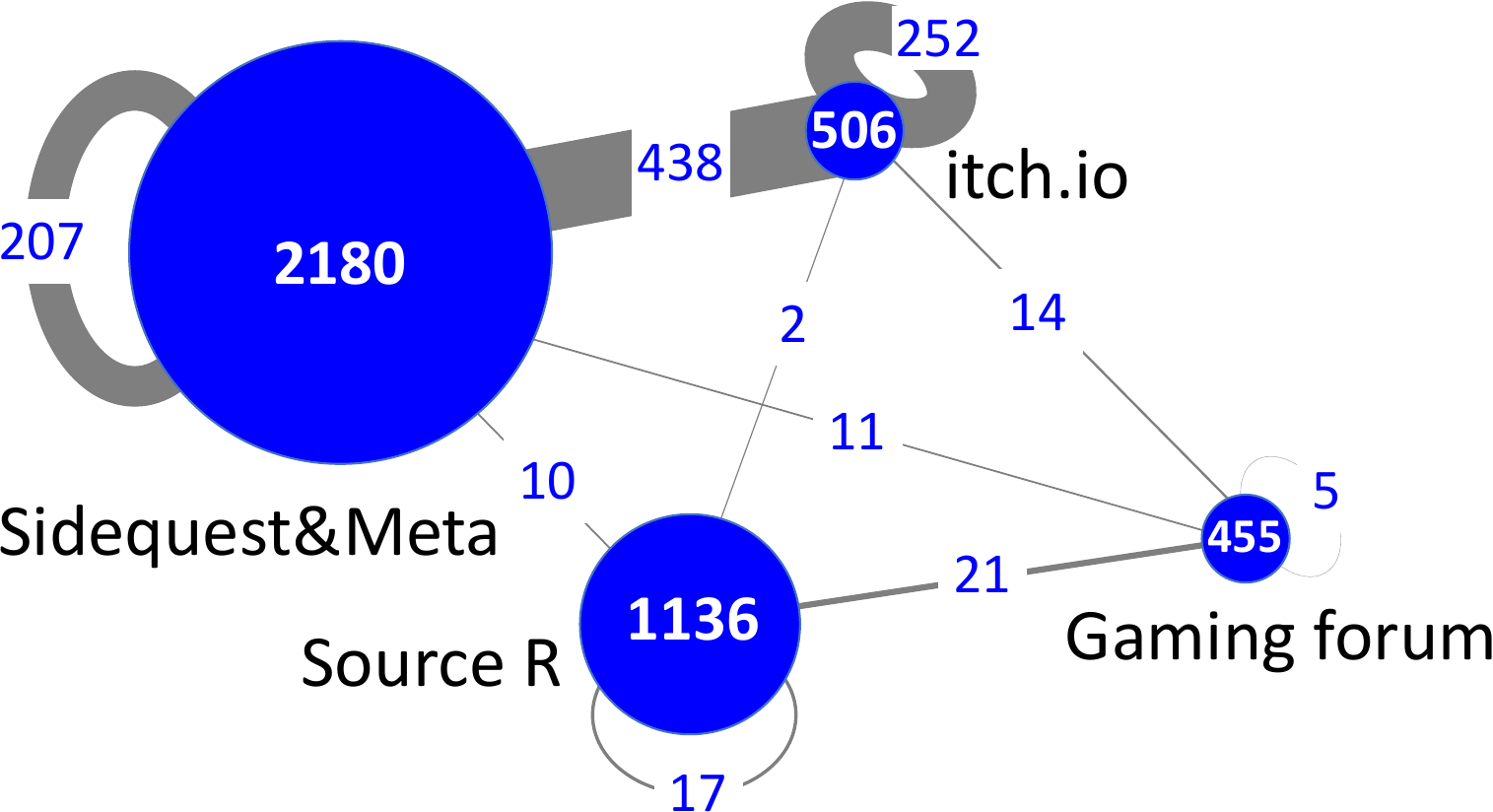}
        \caption{Cloning within and across app sources (node: source; edge: cloning)}
        \label{fig:crossSource}
    \end{minipage}
\end{figure}

\subsection{Fine-grained Detection}
\label{sec:evaluation-fine}
\head{Mesh Decimation Strategy.}
As mentioned in \textsection\ref{sec:fineStage}, we enhance \textsc{3DScan}~\cite{Zuo2023} by applying the \emph{mesh decimation} approach. To evaluate the effectiveness of our improvement, we extracted 8,000 mesh files from 500 randomly selected VR APK files. 
As shown in Fig.~\ref{fig:MeshProcessingTime}, the results indicate that \emph{mesh decimation} reduces average processing time by 53.54\%, with decimation ratios of 0.4 for medium-size meshes ($0.2\,\text{MB} < \text{file size} \leq 1.0\,\text{MB}$) and 0.2 for large-size meshes ($\text{file size} > 1.0\,\text{MB}$). 
This substantial reduction in processing time highlights the efficiency gains introduced by the \emph{mesh decimation} strategy.

The performance improvement stems from the simplified geometric complexity of the models by \emph{mesh decimation} while preserving their essential structure. Importantly, two identical meshes remain identical even after the decimation process, ensuring that the technique does not introduce false negatives. To evaluate the potential for false positives (i.e., a large-sized mesh file can potentially be the same as an unsimplified smaller mesh after mesh decimation), we processed 3,000 mesh files and computed their hash values. No false positives were observed during this evaluation. This is primarily because 
the {mesh decimation} process distorts (even overlaps) the model's geometric details, which would not occur in unsimplified meshes. As a result, they can effectively be distinguished.

\head{Determining Threshold.}
To determine the threshold for pairwise similarity calculation in the fine-grained stage, we use the subset of 100 VR APKs with known clone truths again (as described in \textsection\ref{sec:evaluation-coarse}). After calculating pairwise similarities, we evaluate accuracy across a range of similarity thresholds. By analyzing the false positive and false negative rates at different thresholds, we identify 80\% as the optimal threshold. A detailed accuracy analysis is in \textsection\ref{sec:accuracyAnalysis}.

Fig.~\ref{fig:NodeCounts} illustrates the distribution of node counts in VR app scene hierarchy trees, where the majority of trees contain fewer than 1,500 nodes. For trees with exceptionally large sizes, we apply a tree pruning strategy to enhance comparison efficiency. Specifically, we remove non-contributory nodes, such as \texttt{\small GameObject} nodes with zero components, since they do not affect the functional characteristics of the scene. This approach reduces the computational overhead and ensures the comparison process remains efficient for large and complex trees.

\head{Fine-grained Detection Results.}
After the coarse-grained stage, we identify 416,385 suspected clone pairs. 
The fine-grained stage ultimately yields 977 suspicious app clone pairs involving 307 distinct apps, representing 7.18\% of the dataset, as detailed in Table~\ref{tab:results}. The fine-grained stage requires a total processing time of 30,957.52 seconds. As shown in Fig.~\ref{fig:crossSource}, cloning is observed not only within specific sources but also across different sources. The analysis of the motivations behind the discovered suspicious cloned applications will be discussed in \textsection\ref{sec:discussion}.

\begin{table}[t]
  \caption{Evaluation results of our two-stage approach (SQ denotes SideQuest)}
  \label{tab:results}
  \centering
  \footnotesize
  \begin{tabular}{lc|cc|cc}
    \toprule
    Source & \# of Apps & Coarse (Results) & Coarse (\%) & Fine (Results) & Fine (\%)\\
    \midrule
    Forum         & 455       & 108     & 23.74\%  & 47      & 10.33\%  \\
    itch.io       & 506       & 197     & 38.93\%  & 96      & 18.97\%  \\
    SQ \& Meta & 2,180     & 617     & 28.30\%  & 115     & 5.28\%   \\
    Source R      & 1,136     & 126     & 11.09\%  & 49      & 4.31\%   \\
    Total         & 4,277     & 1,048   & 24.50\%  & 307     & 7.18\%   \\
    \midrule
    App pairs     & 9,144,226 & 416,385 & 4.5535\% & 977     & 0.0107\% \\
    \bottomrule
  \end{tabular}
\end{table}

\subsection{Cloning Verification and Accuracy Analysis}
\label{sec:accuracyAnalysis}
\head{Cloning Verification.}
The cloning verification process is an essential step to confirm the validity of detected potential clone VR apps and determine the accuracy of our VR-Themis. 
First, we analyze the APK signature information of the identified suspicious clones. Apps signed by the same developer are excluded from being classified as clones, as they may represent different versions released by the same author. After filtering, we proceed to verify the clones by installing, executing, and experiencing these VR apps. By evaluating the app's visual design and game mechanics, we ultimately confirm whether they are indeed clone apps. Notably, a few VR apps crash upon startup. Since the authors do not release a fix, we review them by checking previous gameplay videos.

\head{Accuracy Analysis.}
Unlike Android app clone detection, where datasets and standardized benchmarks (e.g., RePack~\cite{Li2019}) are widely adopted, the VR domain lacks both labeled datasets and reference methods. This absence of established baselines precludes direct numerical comparisons. To overcome this limitation, we evaluate VR-Themis through two complementary strategies:

(i) We created a labeled dataset of 200 apps, comprising 100 VR apps with manually identified clone truths (not used for threshold/parameter determination to avoid data leakage) and 100 apps with our artificially generated clones. To generate cloned apps, we refer to methods published in the game communities~\cite{Roldan2022a,Roldan2022b,Thume2019}, including packaging apps with different script backends, using Ghidra~\cite{Ghidra2019} to decompile and modify assembly code, and translating the apps into different languages. The generated cloned apps are solely for internal research purposes and are not disseminated to the public. Then, we use VR-Themis to detect these 200 labeled apps. The results indicate that we successfully identified all cloned pairs with no false positives or false negatives among them.

(ii) For the VR-Themis detection results of 4,277 apps, we manually reviewed all identified clones and found zero false positives. The examination of false negatives is more challenging. First, we define 10 keywords representing themes (e.g., soccer and museum) and randomly collect 10 APKs for each keyword. After manually reviewing each theme of apps, we did not identify any instances of clones that were undetected, suggesting that false negatives are unlikely.

\begin{figure}[t]
    \centering
    \begin{minipage}[t]{0.48\textwidth}
        \centering
        \includegraphics[width=0.95\textwidth]{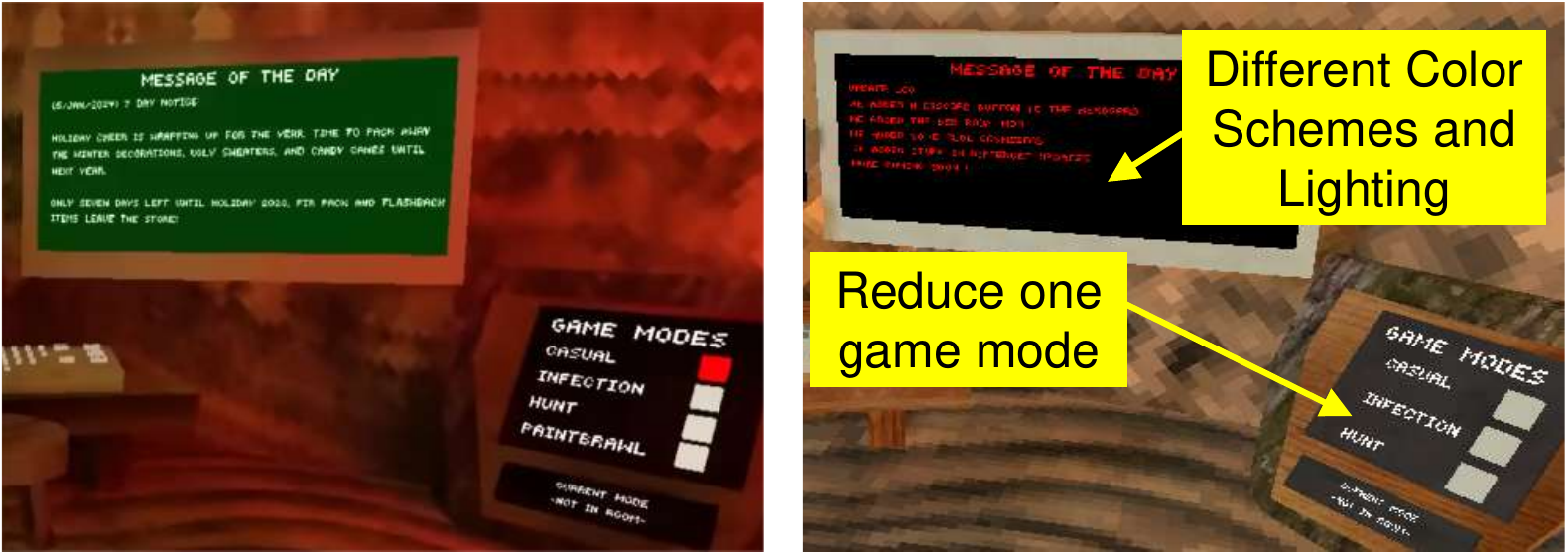}
        \caption{Case 1 for likely unauthorized clones (original on the left, suspected on the right)}
        \label{fig:case1}
    \end{minipage}
    \hfill
    \begin{minipage}[t]{0.48\textwidth}
        \centering
        \includegraphics[width=0.95\textwidth]{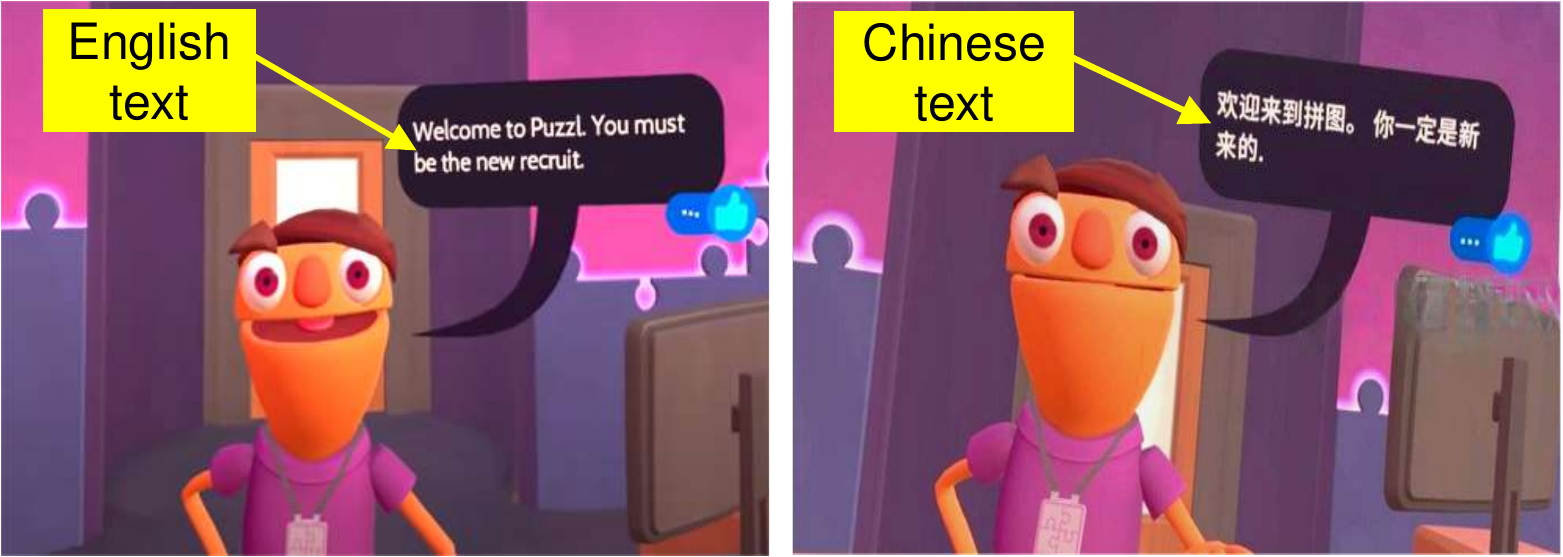}
        \caption{Case 2 for likely unauthorized clones (original on the left, suspected on the right)}
        \label{fig:case2}
    \end{minipage}
\end{figure}

\subsection{Case Studies}
\label{sec:casestudy}
We conduct a detailed analysis of cloning behaviours in two cases. We clarify that our intention is not to act as ``cyber police''. Therefore, we anonymize the app names when presenting them as follows.

\head{Case 1.} 
Cluster 10---the largest cluster identified---contains clone apps suspected to originate from a popular app in the Meta app store (store ID prefix: \texttt{49790}\textasciitilde). These clones are distributed across the SideQuest and itch.io platforms, featuring varied levels of modification to the original app, such as adding \textit{3D} models, altering textures, and introducing new levels. Many of the clones show a similarity of over 80\%, with some exceeding 95\%. An example is illustrated in Fig.~\ref{fig:case1}.

\head{Case 2.} 
App A (MD5 prefix: \texttt{ca23d}\textasciitilde) and App B (MD5 prefix: \texttt{4e59e}\textasciitilde) are different language versions of the same game. Since the original game provides limited language options, attackers repackaged it into an additional language version. Beyond the language change, the clone version also shows evidence of decompilation and adjustments to the hierarchy tree structure during repackaging. Fig.~\ref{fig:case2} shows a screenshot of this cloned case.

\section{Discussion}
\label{sec:discussion}
\head{Handling Shared Assets and Templates.}
To handle cases where different VR apps share common \textit{3D} models, animations, and interaction templates, VR-Themis goes beyond visual similarity by incorporating structural similarity (i.e., HED) and Behavioural similarity (i.e., SBS). This mechanism ensures that apps sharing common assets and templates are not mistakenly flagged as clones under normal circumstances. However, in some edge cases---particularly with lightweight VR applications that rely heavily on publicly available assets or templates---the boundaries between legitimate reuse and cloning may become less clear. Although such cases were not observed in our experiments, future improvements could compare assets in VR apps against publicly available asset libraries (e.g., the Unity Asset Store) to further refine VR-Themis.

\head{Motivations Behind Apps Susceptible to Cloning.}
We analyzed all 307 detected cloned apps and identified three common motivations.
(1) Popular gaming applications: High-profile paid games are prime targets due to their popularity and high-quality \textit{3D} assets. For example, \textit{Gorilla Tag} was cloned across multiple platforms with modified models and scenes.
(2) Language-specific applications: Apps with limited language support are frequently cloned and localized; several English apps were repackaged into Chinese versions distributed through unofficial channels.
(3) Freemium applications: Apps offering free trials are commonly cloned into pirated versions with all paid features unlocked.

\head{Limitations.}
The main limitation lies in clone traceability: determining which app is the original among detected pairs is challenging, and heuristic solutions such as examining submission times and download counts are susceptible to attacks. Additionally, runtime-loaded assets and Unity Asset Bundles may result in a partially captured scene hierarchy, though the latter has minimal impact since most VR apps operate offline. 
Moreover, some applications use code obfuscation for anti-cheating purposes, which can prevent full decompilation. Nevertheless, we find only two such instances among our evaluations, indicating that it has a minimal impact.
In terms of the range of app collections, we are unable to collect paid-download apps and those with broken download links.
Additionally, VR apps may be developed by game engines other than Unity, such as Unreal and libGDX. Moreover, they may be deployed on other VR devices like HTC VIVE Pro 2, Sony PlayStation VR 2, and Pico 4, though many of them also use Android-based operating systems (which are also supported by our tool). In the future, we plan to conduct clone detection research on VR apps developed on more platforms and other engines.

\section{Related Work}
\head{Mobile App Clone Detection.}
Research on repackaged mobile apps primarily focused on the Android platform. 
For instance, DNADroid~\cite{DNADroid} compares code dependence graphs, while DroidMOSS~\cite{DroidMOSS} employs fuzzy hashing. FSquaDRA~\cite{FSquaDRA} utilizes resources to perform similarity comparisons, and DroidEagle~\cite{DroidEagle} focuses on UI layout comparisons. Additionally, methods such as ResDroid~\cite{ResDroid} and ViewDroid~\cite{ViewDroid} integrate both resource and layout comparisons. PiggyApp~\cite{Zhou2013}, another notable approach, constructs vectors using normalized values derived from extracted features.
To the best of our knowledge, there has been a lack of studies specifically addressing the detection of black-box VR apps. 
Existing works, such as those by Chen et al.~\cite{chen2025unveiling} and Huang et al.~\cite{huang2024study}, have primarily focused on code clone detection in open-source VR software. However, these approaches are not applicable to closed-source VR applications, where access to the source code is unavailable. In contrast, our work emphasizes clone detection in closed-source VR APKs, extending the scope beyond code to include a comprehensive Hierarchy-Object-Behaviour structure of VR applications.

\head{Studies on VR/AR Apps.}
Previous works have conducted several studies to understand VR/AR (XR) apps from different perspectives. 
Rodriguez and Wang~\cite{Rodriguez2017} conducted an empirical study on 1,156 open-source VR projects, noting their steady growth, game focus, and frequent mis-committing of auto-generated files.
Li et al.~\cite{LiS2020} conducted an empirical study on 368 real bugs from 33 GitHub-hosted WebXR projects, establishing a bug taxonomy based on symptoms and root causes.
Adams et al.~\cite{Adams2018} focused on VR security and privacy perceptions, conducting a mixed-methods study including semi-structured interviews with 20 VR users and developers, a survey of VR privacy policies, and an ethics co-design study with VR developers.
Guo et al.~\cite{Guo2024} conducted a combination of app analysis, taint analysis, and privacy-policy analysis methods on VR apps to assess security vulnerabilities, privacy data leaks, and contradictory statements in privacy policies. 
Li et al.~\cite{LiS2024} presented Orienter, a zero-shot context-sensitive framework for detecting interactable GUI elements in VR apps.
Zhu et al.~\cite{zhu2025vrexplorer} proposed VRExplorer, a model-based approach for semi-automated testing of VR scenes.
Wu et al.~\cite{wu2026xrfix} explored the use of large language models to repair performance bugs in extended reality applications.
Xu et al.~\cite{xu2026programming} synthesized batch-editing programs of collision meshes in \textit{3D} software from user-provided examples, targeting the same category of non-code assets that dominates VR development.

\head{Unity-based Application.}
Unity game engine is one of the most popular VR application development frameworks. 
We review related studies on Unity. Shim et al.~\cite{Shim2018} performed reverse engineering on Unity-based apps by combining static and dynamic analysis methods to detect malicious apps. 
Zuo et al.~\cite{Zuo2022} designed and implemented a static tool called PaymentScope to automatically identify vulnerabilities in in-app purchasing (IAP) within Unity-based mobile games. Additionally, Zuo et al.~\cite{Zuo2023} proposed a \textit{3D} model clone detection tool named \textsc{3DScan} in mobile games developed with Unity.

\section{Conclusion}
VR app cloning threatens developer interests and user security, yet existing mobile clone detectors fail to capture VR-specific features. In this paper, we presented VR-Themis, a two-stage \emph{Hierarchy-Object-Behaviour}-driven framework that clusters apps via coarse-grained statistical features and performs fine-grained HOB metric comparisons. Experiments on 4,277 VR apps detected 307 clone apps with no false positives, demonstrating its effectiveness and scalability.

\begin{credits}
\subsubsection{\ackname}
This work was supported in part by Hong Kong RGC Project (with Grant No. 12200426) and Grant Seed Funding for Collaborative Research Grants of HKBU (with Grant No. RC-SFCRG/23-24/R2/SCI/06).
We thank the anonymous reviewers for their valuable comments and constructive suggestions.

\subsubsection{\discintname}
The authors have no competing interests to declare that are relevant to the content of this article.
\end{credits}

%
%
%
\bibliographystyle{splncs04}
\bibliography{myreference}
\end{document}